\documentclass[prb,preprint,showpacs,groupedaddress]{revtex4}
\usepackage{dcolumn}
\usepackage{graphicx}
\usepackage{amsmath}
\usepackage{bm}
\usepackage[usenames]{color}
\begin{document}
\title {Quantum Electrodynamics of Quantum Dot-Metal Nanoparticles Molecules}
\author{A. Ridolfo$^1$, O. Di Stefano$^1$, N. Fina$^1$,R. Saija$^1$, and S. Savasta$^1$}
\affiliation{$^1$Dipartimento di Fisica della Materia e Ingegneria Elettronica, Universit\`{a} di Messina Salita Sperone 31, I-98166 Messina, Italy} 
\date{\today}
\begin{abstract}
{We study theoretically the quantum optical properties of hybrid molecules composed of an individual quantum dot and a metallic nanoparticle. We calculate the resonance fluorescence of this hybrid system. Its incoherent part, the one arising from nonlinear quantum processes, results to be enhanced by more than two orders of magnitude as compared to that in the absence of the metallic nanoparticle. 
Scattering spectra at different excitation powers and nonperturbative calculations of intensity-field  correlation functions show that this system can act as  a  nonlinear ultra-compact two-photon switch for incident photons, where the presence (or absence) of a single incident photon field is sufficient to allow (or prevent) the scattering  of subsequent photons. We also find that a small frequency shift of the incident light field may cause changes in the  intensity field correlation function of orders of magnitude.}
\end{abstract}

\maketitle
\section{Introduction}
Control over the interaction between single photons and individual optical emitters deserves great importance in quantum science and quantum engineering \cite{Monroe}. Recently, substantial advances towards the realization of solid state quantum optical devices have been made coupling single quantum dots (QDs) to high-finesse optical cavities \cite{Imamoglu,Fushman}.
An inherent limitation of cavity quantum electrodynamics (QED)  is that the size of the cavity is at least half wavelength and practically much more than that owing to the presence of mirrors or of a surrounding photonic crystal.
Unlike optical microcavities, metallic nanoparticles and metallic nanostructures are able to focus electromagnetic waves to spots much smaller than a wavelength. In this way it is possible to increase the local density of the electromagnetic modes as in microcavities but with ultra-compact structures.
The ability of metallic nanoparticles and nanostructures to control the radiative decay rate of emitters placed in their near field has been widely demonstrated \cite{Sandoghar,Yongdong,Kinkhabwala}. This ability stems from the existence of collective, wave-like motions of free electrons on a metal surface termed surface plasmons (SP) \cite{Maier}.  
Moreover the plasmon excitation covers a broad bandwidth and requires no special tuning to achieve resonance.
An outstanding demonstration of the cavity-like behavior of metallic nanoparticles is the recent realization of a nanolaser based on surface plasmon amplification by stimulated emission of radiation (spaser) \cite{Spaser}.

Optical nonlinearities enable photon-photon interaction and lie at the heart of several proposals for quantum information processing\cite{Monroe,Fushman, Turchette}, quantum nondemolition measurements of photons \cite{Poizat, Nogues}, and single-photon switching \cite{Birnbaum} and transistors \cite{Chang2}.
Recently, the nonlinear optical response of a semiconductor QD coupled with a metallic nanoparticle (MNP) has been theoretically investigated \cite{Bryant1,Bryant2} by exploiting a semiclassical approach where the quantum emitter is treated quantum mechanically and the light field classically. Although this approach provides useful information on the absorption and elastic scattering of this system, the nonlinear optical properties of individual QDs display important quantum optical effects \cite{Imamoglu,Flagg, Akopian, Stevenson}. Moreover, in order to investigate the possible use of these hybrid artificial molecules as quantum devices for the control of individual light quanta, a full quantum mechanical description is required. 
Recently the efficient coupling between an individual optical emitter and propagating SPs confined to a conducting nanowire  has been studied both theoretically \cite{Chang} and experimentally \cite{Akimov}. Non-classical photon
correlations between the emission from the quantum dot and the ends of the nanowire demonstrate that the latter stems from the
generation of single, quantized plasmons.
The potential of this system as  an ultracompact single-photon transistor has been theoretically demonstrated \cite{Chang2}.

Here we investigate the quantum optical properties of a QD-MNP hybrid artificial molecule (see Fig.\ 1a). 
We consider a spherical QD interacting with a spherical MNP of radius $r_{\rm m}$, separated by a distance R (as shown in Fig. 1a). Throughout the paper we consider the exciting input field polarized along the molecule axis (see Fig.\ 1a).
The quantum dot is modeled as a two level system, that is a good approximation when investigating optical processes at frequencies resonant with the lowest energy excitonic transition \cite{Flagg}. Calculations are carried out nonperturbatively, i.e. the QD-MNP interaction as well as quantum fluctuations are treated at all orders. 
We start analyzing the scattering properties of the system as a function of incident power.
We also investigate the modification of the QD resonance fluorescence induced by the presence of the metallic nanoparticle.
Resonance fluorescence of individual quantum emitters provides an interesting manifestation of the quantum theory of light \cite{Walls, Mandel, Scully}. The spectral and quantum statistical properties of the fluorescent light radiated by atom emitters have been the subject of quantum optical measurements for many years. More recently resonant fluorescence of individual QD has been experimentally investigated \cite{Flagg, Ates}. 
We find that the inelastic part of resonance fluorescence, the part arising from nonlinear quantum processes, increases by more than two orders of magnitude with respect to the free-dot value.
We also investigate the statistics of the photons scattered by this system. 
We show that a small variation of the excitation frequency determines a variation of the  second-order correlation function for scattered light beyond three orders of magnitude. These striking results demonstrate that the arrival of one photon is able to suppress or greatly enhance the scattering properties of the second one. 
\section{Quantum description of the metallic nanoparticle-quantum dot interaction}
We consider a spherical QD interacting with a spherical MNP of radius $r_{\rm m}$ separated by a distance $R$.
The system setup is described in Fig.1a. The entire system is embedded in a dielectric medium with constant permittivity $\epsilon_{\rm b}$. The QD dot is described as a two level quantum emitter with a dipole moment $\mu$. This artificial molecule is excited by an applied electromagnetic field $E_{i} = E_0 e^{-i \omega_i t} + c.c.$ polarized along the system axis.
The positive frequency component of the electric field oscillating as $\exp(-i \omega t)$ felt by a quantum emitter, $E_{\rm QD} = E_0 + E_{\rm m}$, is due to the superposition of the free field input field $E_0$ and the field arising from the induced polarization of the MNP \cite{Bryant1}:
\begin{equation}\label{EM}
	E_{\rm m} =  \frac{1}{4 \pi \epsilon_{\rm 0}\epsilon_{\rm b}} \frac{s_{\alpha}P_{\rm m}}{R^{3}}\, ,
\end{equation}
where $s_{\alpha} = 2$ for an applied electrical field  parallel to ${\bf R}$ ($s_{\alpha} = -1$ for a field orthogonal to ${\bf R}$). $P_{\rm m}$ is the polarization of the MNP,
 \begin{equation}\label{PMNP}
	P_{\rm m} = 4 \pi \epsilon_{\rm 0}\epsilon_{\rm b} \beta r^{3}_{\rm m} \left( E_{0} + \frac{1}{4 \pi \epsilon_{0} \epsilon_{\rm b}} \frac{s_{\alpha}P_{\rm x}}{R^{3}} \right)\, .
\end{equation}
$P_{\rm x} = \mu \langle \sigma \rangle$  is the positive frequency component of the QD polarization, being $\langle \sigma \rangle$ the expectation value of the lowering transition operator $\sigma = \left| g \rangle \langle e \right|$. 
The frequency dependent complex coefficient $\beta = (\epsilon_{\rm m}( \omega)-\epsilon_{\rm b})/(2 \epsilon_{\rm b} + \epsilon_{\rm m} ( \omega ))$ determines the SP dipole resonance frequency $\omega_{\rm sp}$ satisfying  $\text{Re} [\epsilon_{\rm m}(\omega_{\rm sp})]=-2\epsilon_{\rm b}$. Performing the first order expansion of  $\text{Re} [\epsilon_{\rm m}(\omega)]$ around $\omega_{\rm sp}$, 
$\beta$ can be well approximated by the complex Lorenzian  
\begin{equation}
	\beta   \cong \,\  \dfrac{3 i \epsilon_{\rm b}\eta}{i\left( \omega_{\rm sp}-\omega \right) + \dfrac{\gamma_{\rm sp}}{2}}\, , 
\end{equation}
with
 \begin{equation}
	\eta = \left(\dfrac{d \,\ \text{Re}[\epsilon_{\rm m}\left(\omega \right)]}{d\omega}\right)^{-1}_{\omega \,\ = \,\ \omega_{\rm sp}}
\end{equation}
and $\gamma_{\rm sp} = 2 \eta\, \text{Im}[ \epsilon_{\rm m}(\omega_{\rm sp})]$.
This approximation enables the  description of  SP resonances within the quasi-mode approach, largely exploited in the framework of cavity QED \cite{Walls, Scully}. Within this scheme, it is possible to describe  the SP field as  single resonant mode with a linewidth arising from the interaction with a reservoir \cite{Bergman}.
The full quantum dynamics of the coupled nanosystem can be derived from the following master equation for the density operator,
\begin{equation}\label{model}
    \dot{\rho} = \frac{i}{\hbar} [\rho, H_{\rm S} ] + {\cal L}_{\rm sp} + {\cal L}_{\rm x}\, .
\end{equation}
The system Hamiltonian is $H_{\rm S} = H_0 + H_{\rm int} + H_{\rm drive}$ with
\begin{equation}
	H_0 = \hbar \omega_{\rm sp}\,  a^\dag a + \hbar \omega_{\rm x}\, \sigma^{\dag} \sigma\, ,
\end{equation}
being $a$ the Bosonic destruction operator describing the SP field mode and $\hbar \omega_{\rm x}$ the energy of the QD excitonic transition.
The Hamiltonian term describing the interaction between the QD exciton and the quantized  SP field, in the rotating wave approximation reads
\begin{equation}
    H_{\rm int} = i \hbar g \left(a^{\dag}\sigma - a \sigma^{\dag} \right)\, ,
\end{equation}
where $\hbar g = \mu {\cal E}$, being $i {\cal E} a = \hat E^+_{\rm m}$ the positive-frequency electric field operator at the QD position.
The system excitation by a classical input field is described by the following Hamiltonian term,
\begin{equation}
    H_{\rm drive} = - E_{\rm 0} \left( \chi a^{\dag} + \chi^{*} a \right) -\mu E_{\rm 0}\left(\sigma^{\dag} + \sigma \right)\, .
\end{equation}
The Markovian interaction  with reservoirs determining  the  decay rates $\gamma_{\rm x}$ and $\gamma_{\rm sp}$ for the QD exciton and the SP mode respectively,  is described by the following Liouvillian terms \cite{Scully},  
\begin{equation}
    {\cal L}_{\rm sp} = \frac{\gamma_{\rm sp}}{2} (2 a \rho a^\dag - a^\dag a \rho - \rho a^\dag a)\, ,
\end{equation}
\begin{equation}
    {\cal L}_{\rm x} = \frac{\gamma_{\rm x}}{2}(2 \sigma \rho \sigma^{\dag} - \sigma^{\dag}\sigma \rho - \rho \sigma^{\dag}\sigma)\, .
\end{equation}
Starting from \ref{model}, the coupled equations of motion for the SP-field expectation values   $\left<  a \right> \equiv \text{Tr} [a \rho]$ and for the emitter   transition-operator determining the QD polarization $\left<  \sigma \right> \equiv \text{Tr} [\sigma \rho]$ can be obtained,
\begin{equation}\label{aa}
    \dfrac{d}{d t} \langle a \rangle = -\left[ i \left(\omega_{\rm sp}-\omega \right) + \dfrac{\gamma_{\rm sp}}{2} \right]\langle a \rangle + g\langle \sigma \rangle + i \dfrac{\chi E_{\rm 0}}{\hbar}\, ,
\end{equation}
\begin{equation}\label{sig}
    \dfrac{d}{d t} \langle \sigma \rangle = -\left[ i \left(\omega_{\rm x}-\omega\right)+ \dfrac{\gamma_{\rm x}}{2} \right]\langle \sigma \rangle - g\langle a \rangle + 2g\langle a \sigma^{\dag}\sigma \rangle + \dfrac{i \mu E_{\rm 0}}{\hbar}\left( 1 - 2\langle \sigma^{\dag}\sigma \rangle \right)\, .
\end{equation}
The equation of motion \ref{sig} for the exciton operator expectation value is  coupled to higher order expectation values.
It is known in cavity QED that   master equations as (\ref{model}) induce an open hierarchy of dynamical equations which needs some approximation. A widely adopted truncation scheme is the one based on the smallness of the excitation density which allows to truncate with respect to the number of photon number-states to be included \cite{Carmichael}.  The inclusion of one-photon states only gives linear optical Bose-like dynamics. Nonlinear optical effects need the inclusion of multi-photon states.  This truncation scheme provides accurate results also for strongly interacting systems provided that a sufficient number of photon-states are included.
At steady state \ref{aa} can formally be solved as,
\begin{equation}\label{a}
   \langle a \rangle = \dfrac{g}{\left[ i \left(\omega_{\rm sp}-\omega\right)+ \dfrac{\gamma_{\rm sp}}{2} \right]}\langle\sigma\rangle + \dfrac{i \chi E_{\rm 0}}{\hbar\left[ i \left(\omega_{\rm sp}-\omega\right)+ \dfrac{\gamma_{\rm sp}}{2} \right]}\, .
\end{equation}
Equating the obtained electric field expectation value $E_{\rm m} =i {\cal E} \langle a \rangle$ with the corresponding field in  \ref{EM}, we obtain,
\begin{equation}
  {\cal E}= \dfrac{s_{\alpha}}{R^3}\sqrt{\dfrac{3\hbar \eta r^{3}_{\rm m}}{4\pi \epsilon_{\rm 0}}}\, ,
\end{equation}
and 
\begin{equation}
  \chi = \epsilon_{\rm b} \sqrt{12 \pi \hbar\eta\epsilon_{\rm 0}r^{3}_{\rm m}}\, .
\end{equation}
\begin{figure}[hbtp]  
\includegraphics[height=60mm]{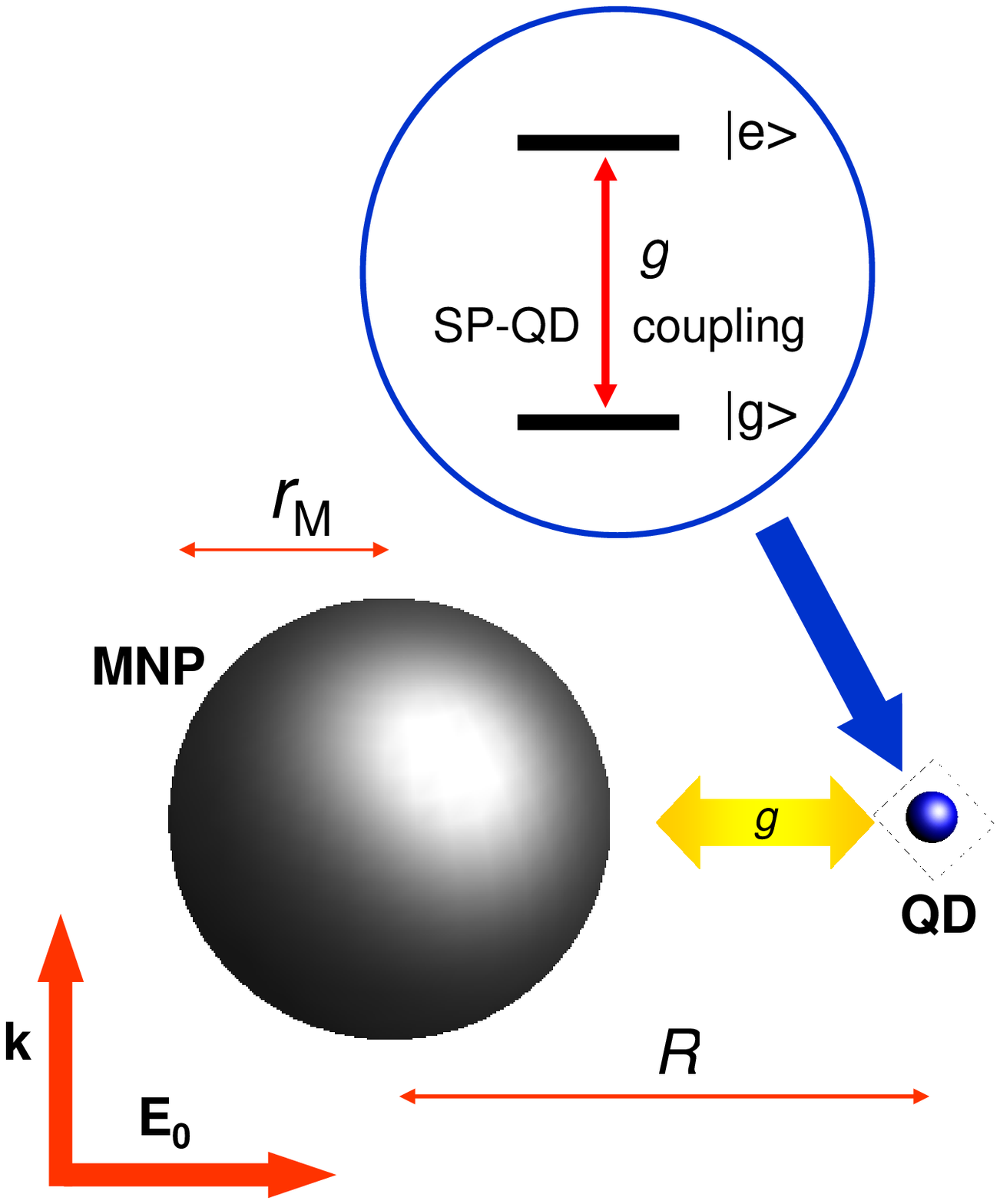}  \hspace{.5 cm} 
\includegraphics[height=80mm]{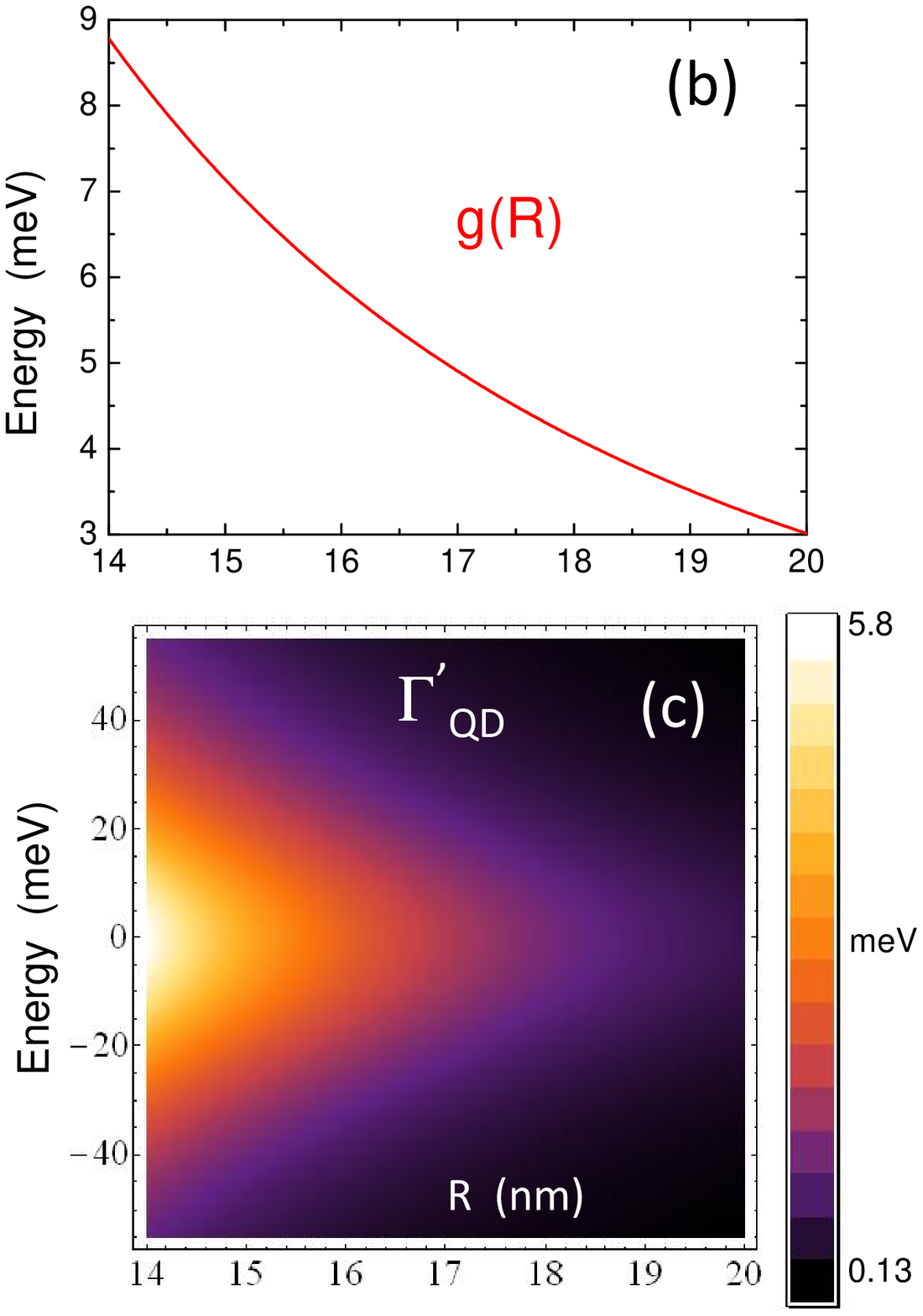}  
\caption{(a) Interaction between a quantum dot and a silver nanosphere: the applied electromagnetic field induces a polarization that causes dipole-dipole coupling. States $\left| g \right>$ and $\left| e \right>$ are coupled via the localized surface plasmon dipole mode with a strength $g$. (b) Dependence of the coupling $g$ on the metallic nanoparticle-quantum dot distance . (c) Contour plot of the effective damping rate of the quantum dot due to the presence of the metallic nanoparticle as a function of distance and detuning.}
\end{figure}
Figure 1b displays the dependence of $g$ on the QD-MNP distance $R$. Throughout the paper we use a dipole moment $\mu = e r_0$  with $r_0 = 0.7$ nm (corresponding to 33.62 Debye), being $e$ the electron charge. 
Throughout the paper we use $\epsilon_{\rm b}=3$ and we consider a silver MNP whose frequency-dependent dielectric permittivity is taken from Ref. \cite{JC}.
Inserting \ref{a} into the second term on the r.h.s. of \ref{sig} , we derive the effective damping rate of the QD due to the presence of the MNP:
\begin{equation}
	\Gamma'_{\rm QD} = \frac{g^2}{\hbar^2} \frac{\gamma_{\rm sp}}{(\gamma_{\rm sp}/2)^2 + (\omega_{\rm sp} - \omega)^2 }\, .
\end{equation}
Fig.\ 1c shows the dependence of $	\Gamma'_{\rm QD}$ on the QD-MNP distance and energy detuning.
Finally comparing the expression for the MNP polarization  \ref{PMNP} with \ref{a}, we obtain 
\begin{equation}
	P_{\rm m} = \chi \langle \ a \rangle\, .
\end{equation}
After the determination of ${\cal E}$ and  $\chi$, \ref{model}  establishes a precise theoretical framework for nonperturbative SP-QED.
In this way the dynamics of the emitter interacting with the field mode can be treated beyond the Wigner-Weisskopf approximation \cite{Chang2}.
\section{The MNP-QD Molecule as a Saturable Scatterer}
We calculate the Rayleigh scattering adopting the standard method  which is valid when the size of the scattering objects is much smaller than the wavelength of incident light \cite{Bryant1}. In this case the scattering intensity is proportional to 
$I_s =\langle\hat P^{-} \hat P^{+} \rangle$, where $	\hat P^{+} = \chi  a  + \mu \sigma$ is the total polarization operator and $\hat P^{-} = (\hat P^{+})^\dag$. It is worth noticing that the scattered intensity contains a coherent part $I^{\rm coh}_{s} = \left| \langle \hat P^{+} \rangle \right|^2$ as well as incoherent contributions with the frequency of the scattered photons not necessarily coincident with that of incident light $I^{\rm incoh}_{s} = \langle \hat P^{-}, \hat P^{+} \rangle \equiv I_s - I^{\rm coh}_{s}$.
\begin{figure}[hbtp]  
\includegraphics[height=100mm]{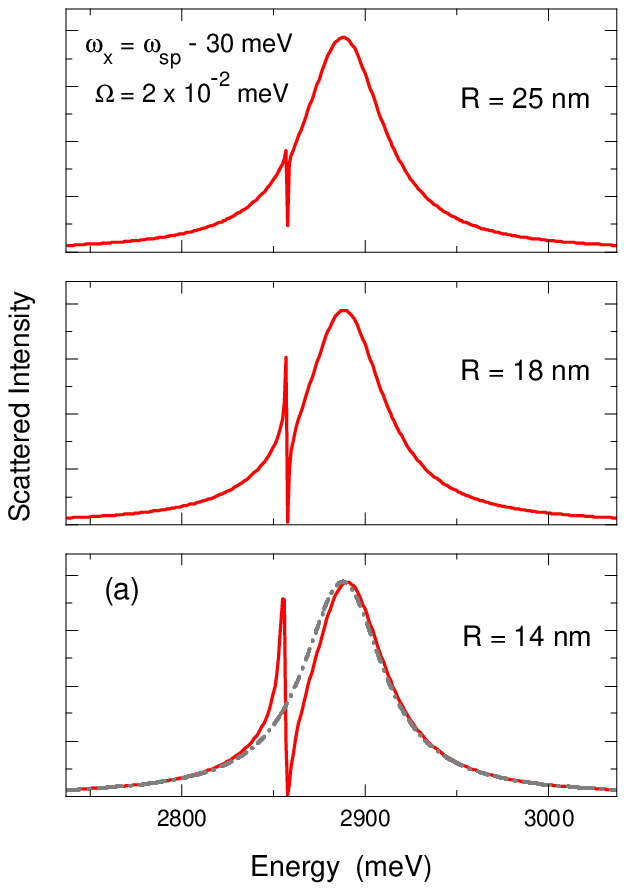}  \hspace{.5 cm} 
\includegraphics[height=100mm]{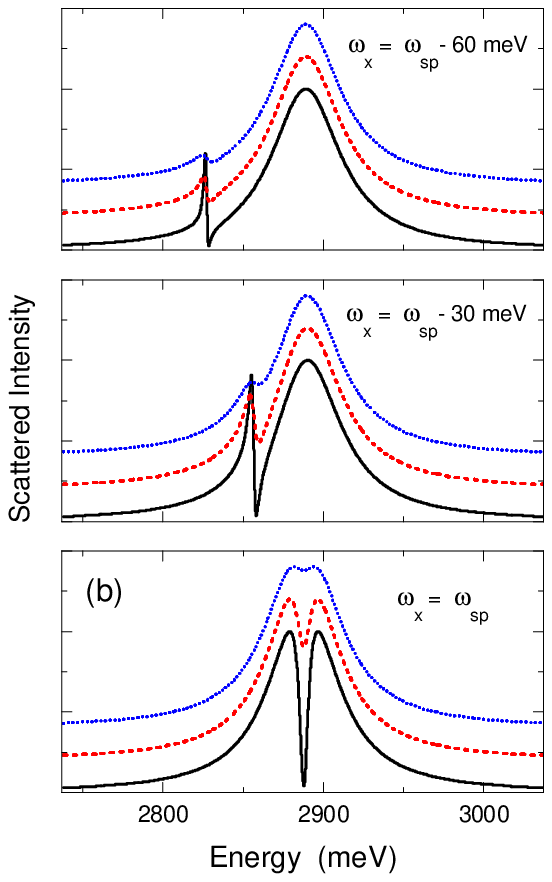}  
\caption{(a) (Color online) Scattered light intensity spectra (red continuous line) calculated for different QD-MNP distances $R$ at low density excitation power. For $R = 14$ nm, the scattered light without the presence of the QD (dot-dashed line) is also plotted. (b) Scattered light intensity spectra calculated at $R = 14$ nm. Each panel shows calculations for a specific exciton-SP energy detuning indicated in the figure. The black continuous line describes plots obtained for an input intensity field $\Omega = 0.02$ meV, the red short-dashed line plots obtained at  $\Omega = 0.4$ meV, and the blue short-dotted line plots at  $\Omega = 1$ meV.}
\end{figure}
Figure 2a displays scattering spectra as function of the frequency of the incidence light  obtained for different QD-MNP distances $R$ as indicated in the panel. The spectra in Fig.\ 2a have been calculated in the limit of very low excitation intensity, where the excitonic populations $\langle \sigma^\dag \sigma \rangle \ll 1$. At $R = 14$ nm a Fano-like lineshape around the QD transition energy $\omega_0$ is evident. For a particular input frequency the scattered light is higly suppressed, while at slightly lower energy an enhancement of scattering due to constructive interference can be observed. For comparison the plot at $R = 14$ nm shows the scattering spectrum in the absence of the QD (dash-dotted line). 
Increasing the distance $R$, the Fano resonance narrows, due to the reduction of $\Gamma'_{\rm QD}$. While at $R = 18$ nm the destructive interference remains almost complete, at larger distances ($R = 25$ nm), the Fano interference effect lowers and the suppression as well as the increase of the scattered light are reduced.
Figure 2b displays scattering spectra obtained for QDs with different excitonic energy levels. 
Interestingly, the interference effect determining a strong suppression of scattering at specific wavelengths of the input field requires no special tuning unlike analogous effects in  cavity QED \cite{Rice}.
While SPs supported by the MNP can be described as harmonic oscillators, the single QD displays nonlinearities at single photon level. Hence we may expect important nonlinear optical effects when increasing the excitation power.
Panel\, 2b puts forward the dependence of light scattering on the intensity of the input field. The continuous lines describe low-field spectra obtained for a Rabi energy $\Omega = 2 \mu E_0 = 2 \times 10^{-2}$ meV. Increasing the input field to $\Omega = 0.4$ meV,  saturation effects appear (dashed line). At $\Omega = 1$ meV saturation is almost complete. The hybrid artificial molecule thus behaves as a frequency dependent saturable  scatterer. 
\section{Resonance Fluorescence}
As outlined  in the previous section, when light excites resonatly a quantum emitter, the output scattered light may be separated into coherent and incoherent parts. The coherent part is due to the elastic Rayleigh scattering and has exactly the same wavelength of the input light: $\omega_{\rm s} = \omega_{\rm i}$. The incoherent part origins from the combination of nonlinear interactions and quantum fluctuations and cannot be described in a semiclassical context . For low incident light intensities elastic scattering dominates, whereas for higher incident light intensities two photon processes with $\omega_{\rm s} + \omega_{\rm s'} = 2\omega_{\rm i}$ take place with the energies of scattered photons $\omega_{\rm s}$ and $\omega_{\rm s'}$ not necessarily equal to $\omega_{\rm i}$. 
We are interested in calculating the steady-state fluorescence emission spectra:
\begin{equation}
  S(\omega_{\rm s}) = \lim_{t \to \infty} 2 \text{Re} \int_0^\infty \left<P^{-}(t), P^{+}(t + \tau) \right> e^{i \omega_{\rm s} \tau} d \tau\, .
\end{equation}
According to the quantum regression theorem \cite{Scully}, two-time correlations $\left< A_n(t) A_m(t + \tau) \right>$ follow the same dynamics of one-body correlation functions $\left<A_m(\tau) \right>$ but with the one-time correlation $\left< A_n(t) A_m(t) \right>$ as initial conditions. 
\begin{figure}[!htbp] \vspace{-0. cm}
\begin{center}
\resizebox{16. cm}{!}{\includegraphics{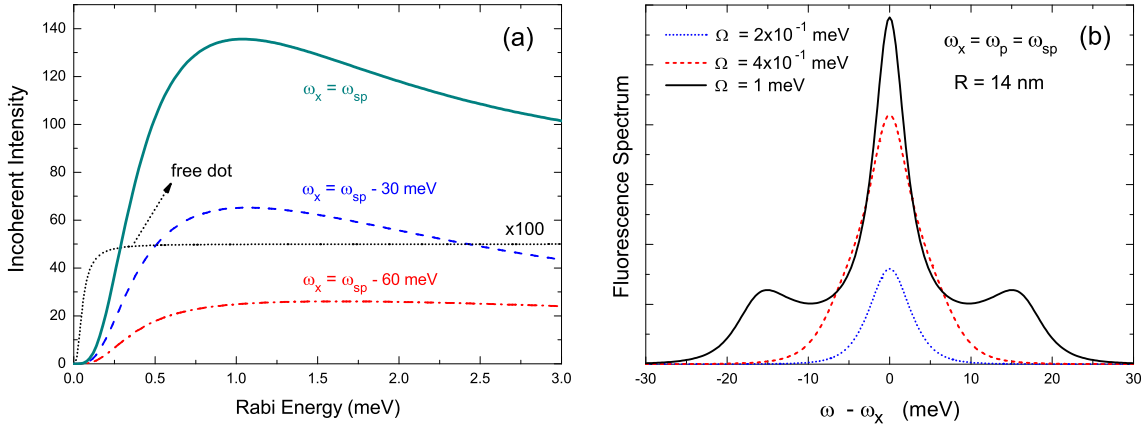}}\caption{ (Color online) (a) Incoherent intensity emission as function of the  driving field intensity obtained for different exciton-SP detunings. The black short-dotted line describes the QD incoherent emission in the absence of the MNP. (b) Resonance fluorescence spectra: intensity of the scattered field as a function of the dected frequency calculated at three different pump intensities.}
\end{center} \vspace{-0. cm}
\end{figure}
The controlled enhancement of single-molecule fluorescence due to its near-field coupling with a metal nanoparticle, acting as a
nanoantenna, has been widely demonstrated. In this section we investigate the strong influence of such a nanoantenna on the nonlinear quantum scattering of an individual quantum emitter. Figure 3a displays the incoherent part of the emitted intensity as a function of the input field for three different excitonic energy levels. Calculations have been performed for  a frequency of the incident field $\omega_i = \omega_{\rm x}$. 
The obtained results show that the MNP is able to strongly influence resonant quantum optical nonlinear processes.
Calculations have been performed considering a QD-MNP distance $R = 14$ nm. At resonance ($\omega_{\rm x} = \omega_{\rm sp}$) we find an enhancement of a factor $\sim 260$ with respect to the QD incoherent emission in the absence of the MNP. The enhancement is the result of two different competing processes: (i) the enhancement of the QD emission due to the SP-increased density of photon modes; (ii) the quenching of the emitted light due to the losses induced by the MNP  described by the rate $\Gamma'_{\rm QD}$ . Fig.\ 3a shows that, despite the MNP enhancement of  the driving field felt by the QD, its presence increases the threshold Rabi energy for the process. This is a direct consequence of the losses increase induced by the coupling of the QD with the MNP. 
Figure 3(b) displays the resonance fluorescence spectra (the incoherent part) as function of the detection frequency for  different excitation powers. It is seen that, with the increasing driving field intensity, the single-peak spectrum around $\omega_{\ rm} = \omega_{\rm x}$ is transformed into the three-peak Mollow spectrum. The emission spectra, owing to the MNP induced losses are strongly broadened with respect  to that of the free QD.
This should enable a better filtering out of the strong elastic component which is necessary in order to measure the Mollow spectrum.
\section{Photon Correlations}
The strongly nonlinear atomic response at the single-photon level leads to pronounced modification of photon statistics that cannot be captured by only considering average intensities, but appears in higher-order correlations of the emitted and scattered fields. Specifically, we focus on the normalized second-order correlation functions $g^{(2)}$ for the scattered field which for a stationary process can be expressed in terms of the total polarization operators as
\begin{equation}
	g^{(2)}(\tau) =\dfrac{\langle \hat P^-(t)  \hat P^- (t+ \tau) \hat P^+(t+ \tau) P^+ (t) \rangle}{\left| \langle  \hat P^- (t)  \hat P^+ (t) \rangle \right|^2}\, .
\end{equation}
\begin{figure}[!htbp] \vspace{-0. cm}
\begin{center}
\resizebox{14. cm}{!}{\includegraphics{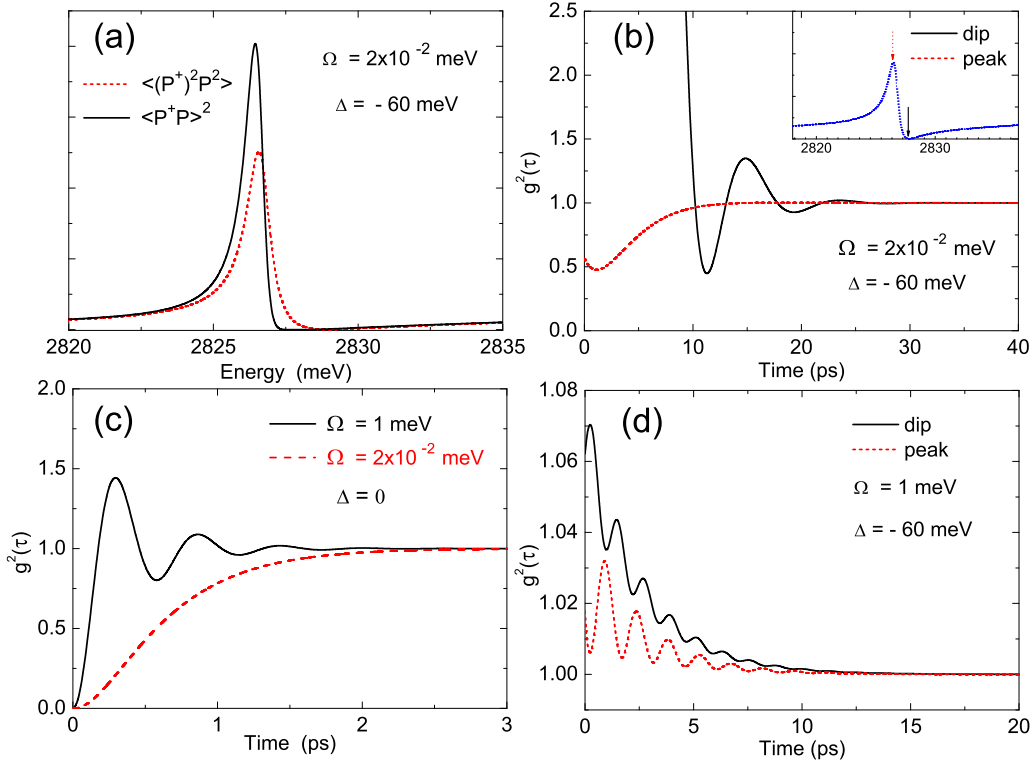}}\caption{ (Color online) (a) Comparison between two photon detection probability (short-dashed red line) and the squared  scattered light intensity (continuous black line) calculated at low excitation power. (b) Normalized second-order correlation functions for the scattered field calculated at low excitation power for two specific frequencies of the driving field indicated by arrows in the inset. The inset displays a detail of the low-excitation power scattering spectrum obtained for $\Delta = \omega_{\rm x} - \omega_{\rm sp} = - 60$ meV. (c) Normalized second order correlation function  for the incoherent part of the emitted field, at high (black continuous line) and low (red short-dotted line) excitation density. (d) As (b) but at higher excitation density.}
\end{center} \vspace{-0. cm}
\end{figure}
Figure 4b shows the normalized second-order correlation functions for scattered photons calculated at low excitation power for two specific frequencies of the driving field indicated by arrows in the inset. The inset displays a detail of the low-excitation power scattering spectrum obtained for $\omega_{\rm sp} - \omega_{\rm x} = 60$ meV.
The continuous line in Fig.\ 4b shows a huge bunching effect. It describes the behavior  of an efficient single-photon switch. Scattering of single photons at this frequency is highly suppresses due to destructive interference, but the arrival of the first photon saturates the QD transition thus enabling the scattering of the second one. For delay times among the two detection events larger than the exciton decay rate, $g^{(2)}(\tau) \to 1$, which is the standard level for coherent classical light. The one that would be measured in the absence of the QD.
The dashed line in Fig.\ 4b describes the second-order correlation function $g^{(2)}(\tau)$ obtained fixing the frequency of the driving field at the Fano-peak corresponding to constructive interference. In this case $g^{(2)}(\tau)$ drops below the classical level  and displays an antibunching effect. Second order correlation functions of the electromagnetic field below 1 cannot be describe by classically and are a signature of the quantum nature of light.
The MNP-QD system thus is able to affect dramatically the photon-statistics of scattered light.
A small variation of the excitation frequency determines a variation of the   second-order correlation function for scattered light beyond three orders of magnitude. This striking effect can be better understood comparing the expectation values for two detection events (the numerator of $g^{(2)}(0)$) and the squared scattering intensity (the denominator) separately as shown in Fig.\ 4a.
$\langle \hat P^-(t)  \hat P^- (t) \hat P^+(t) P^+ (t) \rangle$  displays a reduction of interference effects with respect to   
$\left| \langle  \hat P^- (t)  \hat P^+ (t) \rangle \right|^2$. Only one of the two detected photons is affected by interference: the first photon saturates the QD transition, causing the absence of any interference effect on the second one. 
The second order correlation function for higher excitation power is displayed in Fig.\ 4d. Increasing the excitation power saturates the QD and photon statistics approach that in the absence of the QD.
Fig.\ 4c displays the second order correlation function only for the incoherent part of the emitted field. This field origins from the QD only and its second order correlation function can be calculated as follows,
 \begin{equation}
	g^{(2)}_{\rm incoh}(\tau) = \langle \sigma^\dag (t) \sigma^\dag (t+ \tau) \sigma(t+ \tau) \sigma (t) \rangle /
	\left| \langle  \sigma^\dag(t)  \sigma (t) \rangle \right|^2\, .
\end{equation}
$g^{(2)}_{\rm incoh}(0)$ drops to zero as in standard resonance fluorescence in the absence of the MNP. After the emission of the first photon, the quantum emitter is in its ground state and it needs time in order to get excited once again. Interestingly the presence of the MNP determines a strong decrease of the excitation time. After $\tau = 2$ ps $g^{(2)}_{\rm incoh}(\tau) \sim 1$. Fig.\ 4c has been obtained for $\omega_i = \omega_x = \omega_{\rm sp}$. 
This strong decrease of the excitation time  can be very useful since it reduces the switching time allowing for ultrafast quantum processing.
\section{Conclusions}
We studied theoretically the quantum optical properties of hybrid molecules composed of an individual quantum dot and a metallic nanoparticle. 
The presented numerical calculations have been carried out including quantum fluctuations at all orders. The influence of the MNP on the QD has been treated nonperturbatively beyond the Weisskopf-Wigner approximation.
We applied the developed theoretical framework to investigate the resonance fluorescence of this hybrid system. Its incoherent part, the one arising from nonlinear quantum processes, results to be enhanced by more than two order of magnitude as compared to that in the absence of the metallic nanoparticle.
Scattering spectra at different excitation powers and calculations of intensity-field  correlation functions show that this system can act as  a  nonlinear two-photon switch for incident photons, where the presence (or absence) of a single incident photon field is sufficient to allow (or prevent) the scattering  of subsequent photons. Our calculations have also shown that a small frequency shift of the incident light field may cause changes in the  intensity field correlation function of orders of magnitude.
The ultracompact hybrid system here investigated is in experimental reach. The intriguing quantum optical nonlinear properties here demonstrated provide indications that these hybrid systems could be used as ultra-compact elements in quantum-information technology, and for single-photon devices.
The nonperturbative method here developed to study the coupling of an individual optical emitter to the quantized SP field, can be extended to study the quantum optical properties of more complex structures as a trimer composed by two MNPs and a QD placed in their gap, displaying the vacuum Rabi splitting \cite{SPRabi}.

\end{document}